\documentclass[9pt,arxiv]{lapreprint}

\usepackage{siunitx}    
\usepackage{pdflscape}  
\usepackage{rotating}   
\usepackage{gensymb}    
\usepackage[misc]{ifsym} 
\usepackage{orcidlink}  
\usepackage{listings}   
\usepackage{colortbl}   
\usepackage{tabularx}   
\usepackage{longtable}  
\usepackage{subcaption}
\usepackage{multirow}
\usepackage{csquotes}   

\DeclareSIUnit\Molar{M}

\usepackage[			
	backend=biber,      
    style=phys,   
	natbib=true,		
	hyperref=true,	    
	alldates=year,      
    uniquename=false,   
    maxbibnames=199,     
]{biblatex}

\AtEveryBibitem{
    \clearfield{urlyear}
    \clearfield{urlmonth}
    \clearlist{language}
}

\title{Multiscale Biomolecular Simulations in the Exascale Era}

\author[ \orcidlink{0000-0003-1588-338X} 1]{David Carrasco-Busturia}
\author[ \orcidlink{0000-0001-5513-8056} 2]{Emiliano Ippoliti}
\author[ \orcidlink{0000-0002-3925-3799} 3]{Simone Meloni}
\author[ \orcidlink{0000-0002-1704-8591} 4]{Ursula Rothlisberger}
\author[ \orcidlink{0000-0001-7487-944X} 1 \Letter]{Jógvan Magnus Haugaard Olsen}

\affil[1]{DTU Chemistry, Technical University of Denmark (DTU), DK-2800 Kongens Lyngby, Denmark}
\affil[2]{Computational Biomedicine, Institute for Neuroscience and Medicine INM-9, Forschungszentrum Jülich GmbH, DE-52428 Jülich, Germany}
\affil[3]{Dipartimento di Scienze Chimiche, Farmaceutiche ed Agrarie (DOCPAS), Università degli Studi di Ferrara (Unife), I-44121, Ferrara, Italy}
\affil[4]{Laboratory of Computational Chemistry and Biochemistry, Institute of Chemical Sciences and Engineering, École Polytechnique Fédérale de Lausanne (EPFL), CH-1015 Lausanne, Switzerland}

\metadata[]{\Letter\hspace{.5ex} For correspondence}{jmho@kemi.dtu.dk (JMHO)}
\metadata[]{Funding}{EI thanks the Helmholtz European Partnering program (“Innovative high-performance computing approaches for molecular neuromedicine”) for funding. UR gratefully acknowledges funding from the Swiss National Science Foundation via the NCCR MUST and the individual grants 200020-185092 and 200020-219440, as well as computing time from the Swiss National Computing Centre CSCS. JMHO gratefully acknowledges the financial support from VILLUM FONDEN (Grant No.\ VIL29478).}
\metadata[]{Competing interests}{The authors declare no competing interests.}

\leadauthor{Carrasco-Busturia}
\shorttitle{Multiscale Biomolecular Simulations in the Exascale Era}

\begin{document}

\maketitle

\begin{abstract}
The complexity of biological systems and processes, spanning molecular to macroscopic scales, necessitates the use of multiscale simulations to get a comprehensive understanding. Quantum mechanics/molecular mechanics (QM/MM) molecular dynamics (MD) simulations are crucial for capturing processes beyond the reach of classical MD simulations. The advent of exascale computing offers unprecedented opportunities for scientific exploration, not least within life sciences, where simulations are essential to unravel intricate molecular mechanisms underlying biological processes. However, leveraging the immense computational power of exascale computing requires innovative algorithms and software designs. In this context, we discuss the current status and future prospects of multiscale biomolecular simulations on exascale supercomputers with a focus on QM/MM MD. We highlight our own efforts in developing a versatile and high-performance multiscale simulation framework with the aim of efficient utilization of state-of-the-art supercomputers. We showcase its application in uncovering complex biological mechanisms and its potential for leveraging exascale computing.
\end{abstract}

\section{Introduction}

Biological processes extend across wide scales in space and time due to the hierarchical organization of biological matter~\cite{Schaffer2021}. The characteristic dimensions span from the molecular scale of a few ångströms with ultrafast electronic processes on the order of atto- and femtoseconds and rapid chemical reactions that occur within pico- to microseconds, to the macroscopic scale of cells and organs that are visible to the naked eye and where processes extend to seconds and even days and years. The complexity in living organisms largely stems from this hierarchical structure, where a local process may trigger a cascade of events across multiple spatial and temporal scales. Thus, multiscale approaches integrating different resolutions and methodologies are essential for capturing the entire spectrum of biological events~\cite{Palermo_2020_frontiers, Trovato2021}.

The continuous development of multiscale methods in computational structural biology is driven by simultaneous advancements in algorithms, software implementations, and hardware technology that push the boundaries of molecular simulations in terms of accessible time scales and system sizes, capturing biological systems from the atomistic to the cellular level. In particular, quantum mechanics/molecular mechanics (QM/MM) molecular dynamics (MD) simulations have become increasingly important in the last decades for studying processes where a description of electronic degrees of freedom is paramount and thus beyond the capabilities of classical MD based on standard analytical force fields. Examples of such events encompass all types of chemical reactions, including proton-coupled electron transfers~\cite{hammes-schiffer_2015, Nottoli2021}, photochemistry~\cite{Toldo2023}, and the manifold of chemical transformations observed in enzymatic reactions~\cite{Ramos_2017, Current_Op_2022_Kulik}, especially those involving transition metals~\cite{Tzeliou2022}. A detailed characterization of these phenomena, including reactants, transition states, intermediates, and products, as well as the involved relative free energies, reaction rates, and binding affinities in both electronic ground and excited states, also has direct implications for drug design~\cite{vanderKamp2024}. In QM/MM models of biomolecular systems, a smaller part (the QM subsystem), such as the active site of an enzyme or a chromophore embedded in a protein or lipid bilayer, is described at the quantum mechanical level, while the remainder (the MM subsystem) is modeled using molecular mechanics. This multiscale strategy balances a detailed and accurate but computationally costly description of the essential part of a system with a coarser but computationally expedient approach for the much larger MM part. The dynamic aspect of QM/MM MD is crucial to accurately capture the behavior and function of complex proteins~\cite{Brunk_2015, Clemente2023}. However, QM/MM MD simulations have a substantially higher computational cost than classical (i.e., MM) MD simulations, severely limiting the accessible time scales. Typically, for a density functional theory (DFT)-based QM/MM MD simulation with around 100 atoms in the QM subsystem, accessible time scales are limited to a few hundred picoseconds~\cite{Current_Op_2022_Kulik, drug_jcim}.

The advent of exascale computing marks a pivotal moment for all simulation-based scientific fields~\cite{Remmel_CEN, Exascale_Choi}. This remarkable technological achievement was accomplished by connecting thousands of computing nodes through high-speed network interconnects. Each node combines traditional general-purpose central processing units (CPUs) with powerful graphics processing units (GPUs). However, exascale supercomputer architectures also introduce new challenges since programming software applications for these heterogeneous machines requires a judicious decomposition of the computational work to exploit each component of the machines optimally~\cite{DiFelice2023}. Therefore, to fully exploit the computational power of present and future heterogeneous HPC architectures, a high degree of concurrent parallelism is needed, where different parts of the supercomputer work on different subdomains of the computational model, each described within a different theoretical method. Together, exascale computing and the development of novel computational methods and software can enable longer and more accurate simulations on larger and more complex systems. This opens up new opportunities for discovery and innovation in the life and health sciences, potentially revolutionizing areas such as drug design and bioengineering.

The heterogeneous CPU/GPU technology highlighted above is taken to the next level with modular supercomputer architectures that integrate a variety of hardware technologies into interconnected partitions~\cite{carpenter_2022_6090425, suarez_2022_6508394}. A prime example of this is the LUMI supercomputer~\cite{lumi}, along with the upcoming exascale JUPITER supercomputer~\cite{jupiter}, both procured by the European HPC Joint Undertaking (EuroHPC JU). Currently, LUMI consists of two primary \emph{number-crunching} partitions: a general-purpose CPU partition and a high-performance GPU-accelerated partition. Additionally, it features an interactive data analytics partition and an accelerated storage partition. Looking ahead, the integration of prospective technologies like quantum and neuromorphic computing into traditional HPC infrastructure is expected~\cite{carpenter_2022_6090425, suarez_2022_6508394}. Indeed, the procurement of a quantum computing partition for LUMI is already underway~\cite{lumi-q}, and the Jülich Supercomputing Center in Germany is already experimenting with this kind of integration~\cite{JUNIQ}. Modular supercomputer architectures offer substantial benefits, particularly in allowing calculations to run on the most suitable hardware for the specific problem. Moreover, existing software packages that have yet to be optimized for new hardware remain useful for solving important scientific problems. Fully exploiting the capabilities of complex modular supercomputers for exceptionally challenging scientific problems, which require the full computing power of the machine, necessitates a new computational paradigm. Here, we highlight an approach recently introduced in the field of multiscale biomolecular simulations.

Multiscale methods, especially the QM/MM approach, have proven indispensable for exploring complex biological phenomena. Interestingly, QM/MM is not only a robust technique in itself but also offers a route to overcome some of the existing limitations in the development of software for atomistic simulations on modern hybrid computer architectures~\cite{Remmel_CEN}. Beyond QM/MM, multiscale methods can also be extended to incorporate multiple layers ranging from coarse-grained (CG) to continuum models of matter where parts of a system, e.g., membrane regions that are sufficiently distant from an embedded protein of interest, are treated by techniques usually employed for meso- and macroscopic systems~\cite{Jin2022, Guo2022, Bock2023, BorgesArajo2023, Fadda2024}.

To reach the full potential offered by exascale supercomputers, it is imperative that multiscale interfaces scale efficiently, fully leveraging the extensive network of CPUs and GPUs. That is one of the main objectives of MiMiC (multiscale modeling in computational chemistry), a high-performance and versatile framework for multiscale simulations~\cite{olsen2019mimic}. The \emph{program-agnostic} MiMiC framework is designed to combine virtually any QM and MM (or other) program without compromising the computational efficiency and scalability of the simulation. Indeed, MiMiC, although still in its nascent stages, has demonstrated its ability to efficiently scale QM/MM MD simulations across thousands of CPU-based computing nodes~\cite{mimic_jctc_slava, drug_jcim}. Here we will showcase the MiMiC framework for biomolecular simulations, illustrating its potentiality in leveraging state-of-the-art supercomputing resources and its capability to address complex biochemical and biophysical problems.

\section{Current Status of Multiscale Biomolecular Simulation Software}

Multiscale QM/MM capabilities are often implemented by either extending dedicated QM or MM programs with the functionalities of the other~\cite{Laio_jcp, Gaussian_QMMM1, CP2K_QMMM, CHARMM, AMBER12_overview, deMon2k, ORCA_2020, TeraChem} or by creating ad hoc interfaces between stand-alone QM and MM programs~\cite{Isborn2012, TINKTEP, GAUSSIAN_TINKER1, QUICK_AMBER}. Additionally, some programs offer flexible interfaces that facilitate easy coupling with other programs~\cite{AMBER_QM_MM_MD_interface, NAMD_goes_quantum, Terachem_protocl_buffers_2023}. Beyond these, there are integrative frameworks that do not contain inherent QM, MM, or similar functionalities, but instead depend entirely on other programs for these capabilities~\cite{MSCALE, QMMMW, PUPIL_soft_integration, PUPIL_soft_integration2, Cuby, ASE, Weingart2018, ChemShell_QM_MM, ChemShell_QM_MM_redevelopment, LICHEM, LICHEM2, Janus, Molssi_driver, QM3, QMMM2023}.

The interface between QM and MM components can be classified as tightly or loosely coupled, reflecting the degree of interdependence between the components. Loose coupling is preferred for its flexibility and ease of maintenance, enabling straightforward integration of independent programs. This approach supports the quick incorporation of new functionalities and advancements within individual programs without necessitating alterations to other coupled components. Crucially, loose coupling permits independent optimization of each program for peak performance. However, this flexibility can come at the cost of computational efficiency due to slower inter-program communication, particularly when compared to the tight coupling approach that allows for direct in-memory data sharing.

There are three communication mechanisms for exchanging data between programs, namely the file-, library-, and network-based approaches. Most commonly utilized for general interfaces and integrative frameworks is the file-based approach, which does not require modifications to the source code of the interfaced programs, making it the simplest to implement. Here, the main driver program generates input files for the interfaced programs, executes them, and subsequently reads their output files. However, it is notably the least efficient communication mechanism, primarily due to slow disk write/read operations and overhead associated with executing and terminating the interfaced programs. Both drawbacks are avoided in the library-based approach, where the interfaced programs are converted into a library that is linked to the main program, thus enabling efficient in-memory data exchange. The disadvantages are the rather intrusive modifications needed in the source codes of the interfaced programs, and the risk of ending up with a tight coupling focused on a few specific programs. Moreover, a critical aspect of multiscale implementations is parallel scalability and efficiency. The optimal parallel algorithms for QM and MM components, or even among different QM methods, may differ significantly and may not seamlessly integrate within the more tightly coupled environments.

The network-based approach, on the other hand, offers a solution that potentially circumvents the limitations inherent in both file- and library-based methods. It achieves a balance between flexibility and communication efficiency by enabling data exchange over a network. This facilitates seamless communication between programs running on different processors (CPUs and GPUs), across computing nodes, or even among supercomputer partitions. Such an approach retains the benefits of loose coupling, i.e., ease of integration and maintenance, while surpassing file-based approaches in performance through the use of fast network protocols and remote direct memory access (RDMA) technologies. Crucially, it can be implemented so as not to disrupt the normal execution of interfaced programs, thus preserving their performance. However, the network-based approach requires a more sophisticated design and management strategy to reduce latency and enhance throughput. Despite these challenges, with proper implementation, the network-based approach substantially improves the efficiency and scalability of multiscale simulations by facilitating high-speed communication between diverse computational models.

\section{Toward Multiscale Biomolecular Simulations on Exascale Architectures}

For multiscale simulations to run efficiently on state-of-the-art supercomputers, first and foremost, it is imperative that the programs dedicated to a specific methodology, such as QM and MM, are able to take advantage of the strengths of both CPUs and GPUs. Moreover, these programs must be capable of using the vast parallel processing capabilities of exascale architectures. Considerable efforts are underway to push QM- and MM-based programs towards exascale~\cite{Phillips2020, Pll2020, Kowalski2021, das2022dft, Schade2023, Carnimeo2023, Zahariev2023, Gavini2023} with support from EuroHPC JU through its HPC Centres of Excellence~\cite{EuroHPCCoE} and the Exascale Computing Project (ECP) led by the US Department of Energy~\cite{ecp}.

\begin{figure*}[b!]
\centering
\includegraphics[width=0.8\textwidth]{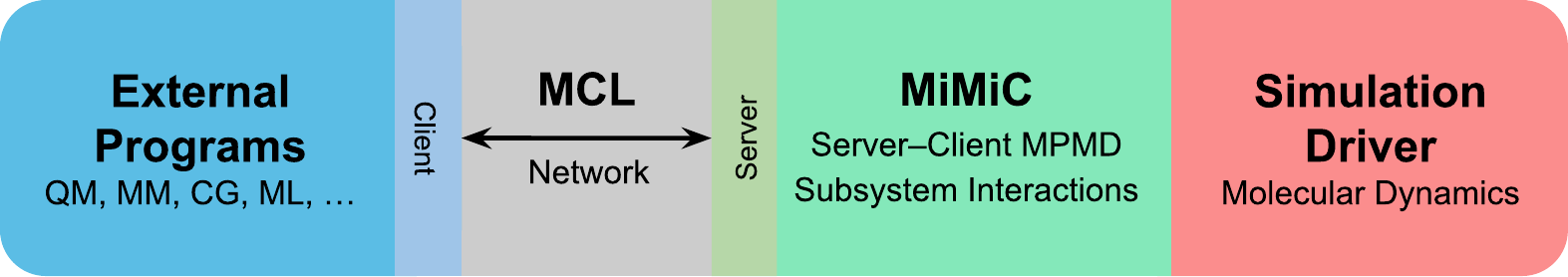}
 \caption{Illustration of the strategy used by the MiMiC framework}
\label{fig:framework}
\end{figure*}

For a multiscale simulation framework to fully exploit the capabilities of these QM and MM programs, it must seamlessly integrate with their optimal parallelization strategies without introducing inefficiencies. This entails enabling the concurrent execution of interfaced programs, eliminating the need for their repeated startup and shutdown, and reducing communication overhead to a minimum. Crucially, the calculation of subsystem interactions, such as those between QM and MM particles, needs to be highly parallel and efficient to prevent it from becoming a computational bottleneck. Implementing effective and automatic load balancing is also critical to ensure that computational resources are utilized optimally, thereby maximizing the throughput of individual simulations.

\begin{figure*}[t!]
\centering
\includegraphics[width=1.0\textwidth]{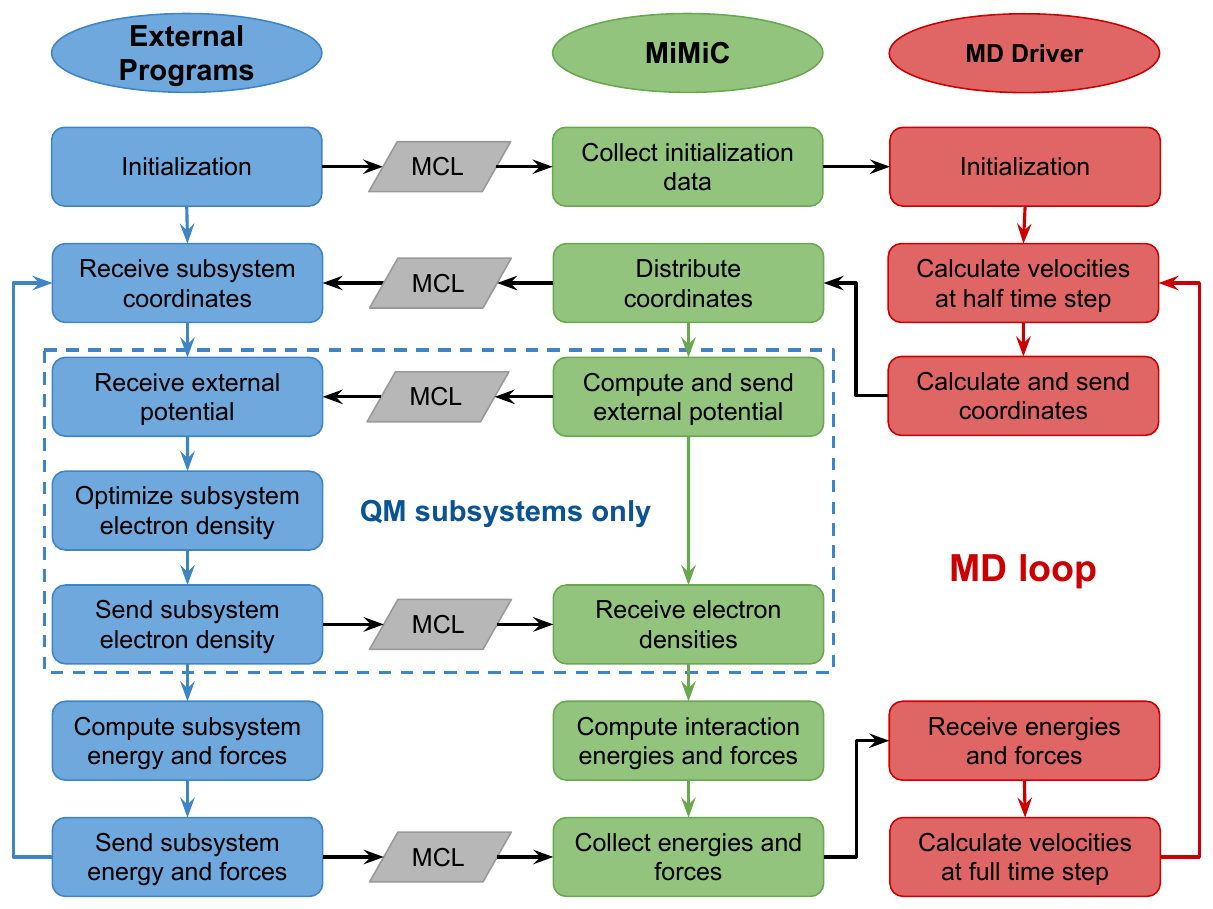}
 \caption{Illustration of MiMiC-based simulation workflow}
\label{fig:workflow}
\end{figure*}

The versatile and high-performance MiMiC framework for multiscale simulations was designed with the aim of addressing the criteria outlined above~\cite{olsen2019mimic}. The strategy used by the MiMiC framework is illustrated in Figure~\ref{fig:framework}. On one side, MiMiC connects to a simulation driver that manages the overall simulation process, including the integration of the equations of motion and the maintenance of temperature and pressure. On the other side, it interfaces multiple external programs using a client-server approach combined with a multiple-program multiple-data (MPMD) model. Each external program is tasked with calculations belonging to a given subsystem, while MiMiC calculates subsystem interactions. Importantly, the external programs run concurrently on separate computational resources using their own optimal parallelization capabilities. Communication between MiMiC and the external programs is facilitated by the MiMiC Communication Library (MCL), a dedicated lightweight library that ensures efficient network-based communication and simplifies the interfacing with external programs. An illustration of a MiMiC-based simulation workflow is shown in Figure~\ref{fig:workflow}. The scope of the MiMiC framework is broad, extending beyond QM/MM. This includes QM/QM, which integrates different levels of QM theory, QM/MM/CG, and even models that incorporate machine-learning (ML) techniques like ML/MM or QM/ML.

\begin{figure*}[t!]
\centering
\includegraphics[width=0.9\textwidth]{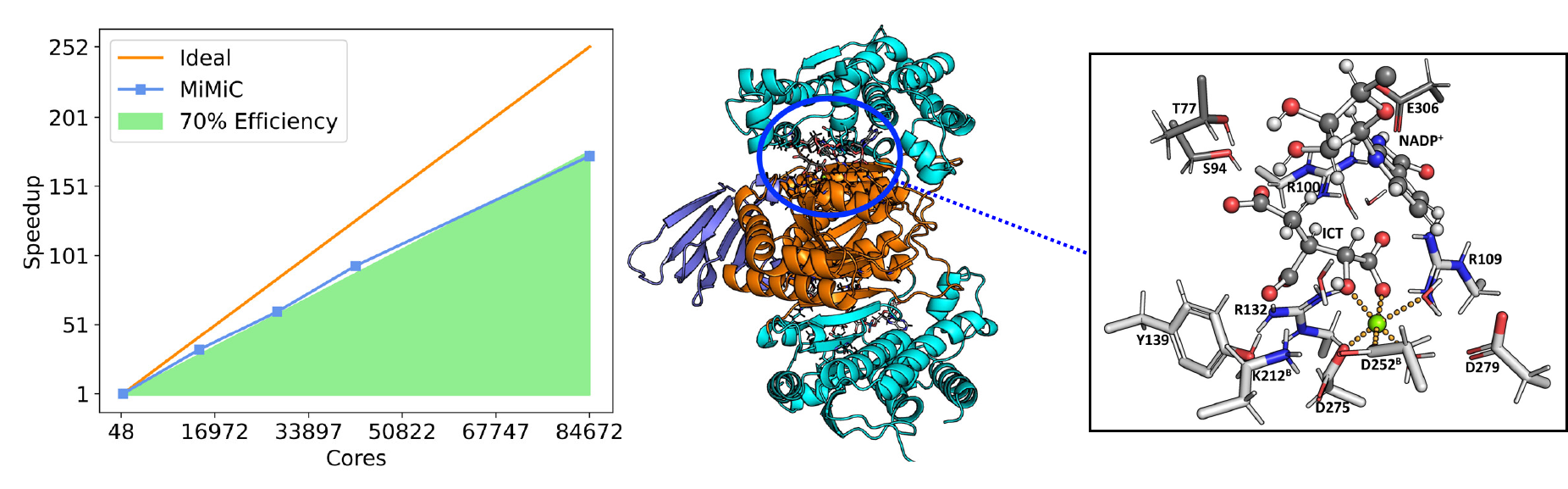}
\caption{\textbf{Parallel scalability and efficiency of MiMiC-based QM/MM MD simulations of IDH1.} The IDH1 system has a total of 130,828 atoms with 142 QM atoms~\cite{drug_jcim}. The simulations were run on the CPU partition of JUWELS~\cite{JUWELS}. The speedup is given in terms of the time per MD step normalized against a reference run using seven nodes. Adapted from Raghavan et al.~\cite{drug_jcim}, licensed under CC BY 4.0.}
\label{fig:cases-scaling}
\end{figure*}

The MiMiC framework has been used to implement a DFT-based electrostatic embedding QM/MM method, employing the CPMD program~\cite{cpmd} as both the MD driver and QM engine, and GROMACS~\cite{Pll2020} as the MM engine~\cite{olsen2019mimic, mimic_jctc_slava}. In electrostatic embedding, the (external) electric field from the point charges in the MM subsystem is included in the Hamiltonian of the QM subsystem, thus directly polarizing its electronic density. The calculation of electrostatic QM/MM interactions is based on a dense grid representation of the electronic density, which is computationally expensive, especially for large systems, because the number of integrals over the electron density scales linearly with the number of atoms in the MM subsystem. The MiMiC framework has implemented an efficient approach that substantially speeds up the calculation essentially without compromising the accuracy of the forces, thus reducing the computational cost by about 80 \% for small systems (e.g., small solvated proteins) and up to 99 \% for larger systems (e.g., membrane-embedded proteins)~\cite{olsen2019mimic}. Furthermore, MiMiC implements a hybrid shared- and distributed-memory (OpenMP/MPI) parallelization strategy to ensure that these calculations remain efficient and do not hinder performance, even under highly parallel conditions~\cite{mimic_jctc_slava}. A powerful example of this is shown in Figure~\ref{fig:cases-scaling}, where the extreme scalability of the CPMD program is harnessed to achieve strong scalability beyond 80,000 CPU cores with a parallel efficiency of 70 \% for a single MiMiC-based QM/MM MD simulation of the human isocitrate dehydrogenase-1 (IDH1)~\cite{drug_jcim}.
Table~\ref{tab:throughput} shows examples of the computational performance and cost of MiMiC-based QM/MM MD simulations using CPMD and GROMACS. It is clear that the simulation throughput depends first and foremost on the chosen exchange-correlation functional and the size of the QM subsystem. This reflects the fact that the QM calculation is by far the most computationally demanding part of a QM/MM MD simulation. Hybrid exchange-correlation functionals are particularly expensive in a plane-wave basis set such as the one employed by the CPMD program. Still, due to the excellent parallelism in CPMD, the simulation throughput can be pushed to 4.8 ps/day for the small QM region (46 atoms) and 0.7 ps/day for the larger QM subsystem (142 atoms). Using instead a non-hybrid functional pushes the performance to 21 and 5.4 ps/day for the small and large QM subsystems, respectively.

\begin{table}[h!]
    \centering
    \caption{\textbf{Computational performance and cost of MiMiC-based QM/MM MD simulations.} All simulations were run with a 0.5 fs timestep on the CPU partition of JUWELS~\cite{JUWELS} at the scaling limit (parallel efficiency $\geq 70 \%$). The systems are p38$\alpha$ mitogen-activated protein kinase (169,550 atoms) and human isocitrate dehydrogenase-1 (130,828 atoms). We refer to the original work for full computational details~\cite{drug_jcim}.}
    \label{tab:throughput}
    \begin{tabular}{lcccc}
        System & \multicolumn{2}{c}{p38$\alpha$} & \multicolumn{2}{c}{IDH1} \\
        QM atoms & \multicolumn{2}{c}{46} & \multicolumn{2}{c}{142} \\
        XC functional & BLYP & B3LYP & BLYP & B3LYP \\
        Throughput (ps/day) & 21 & 4.8 & 5.4 & 0.7 \\
        Cost (node-hours/ps) & 9 & 1280 & 480 & 60480 \\
    \end{tabular}
\end{table}

\section{Biological Applications}

\begin{figure}[ht!]
\centering
\includegraphics[width=0.5\textwidth]{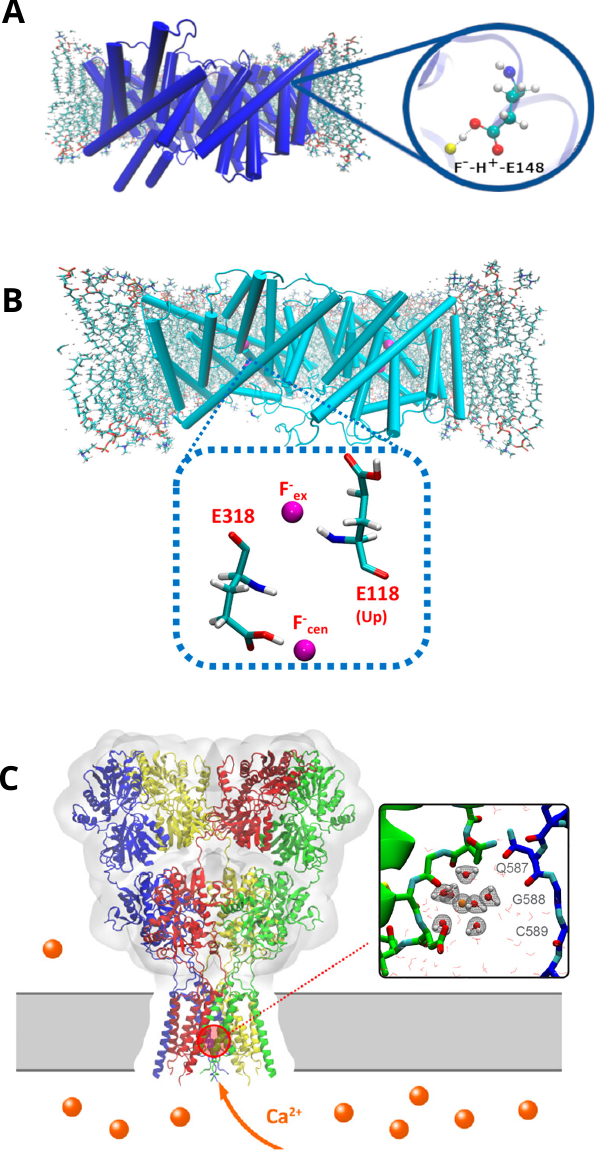}
\caption{\textbf{Illustrations of biological systems that have been studied using the MiMiC framework.}
A: ClC-ec1 anion/proton antiporter embedded in a solvated lipid bilayer with a total of 150,925 atoms (19 QM atoms)~\cite{CLC_jacs}. Reprinted with permission from Ref.~\cite{CLC_jacs}. Copyright 2020 American Chemical Society.
B: CLC\textsuperscript{F}-eca fluoride/proton antiporter embedded in a solvated lipid bilayer with a total of around 174,000 atoms (36 QM atoms)~\cite{CLC-F_jpcl}. Reprinted with permission from Ref.~\cite{CLC-F_jpcl}. Copyright 2021 American Chemical Society.
C: AMPAR cation channel embedded in a solvated lipid bilayer where the transmembrane domain was included in the simulations (22 QM atoms)~\cite{Ca_jcim}. Figure by Schackert et al.~\cite{Ca_jcim}, licensed under CC BY 4.0.
}
\label{fig:cases}
\end{figure}

The MiMiC framework has enabled a number of recent computational studies that demonstrate its utility and efficiency across a broad spectrum of systems with biological relevance. In this section, we showcase select examples that highlight the impactful contributions of MiMiC-based QM/MM MD simulations in advancing our understanding of complex biological processes.

Among the pioneering uses of MiMiC-based QM/MM MD simulations were studies of CLC proteins, a large family of anion channels and transporters. \citeauthor{CLC_jacs} studied the molecular mechanism of fluoride inhibition of the anion/proton exchanger ClC-ec1 from \textit{E. coli} (Fig.~\ref{fig:cases}A)~\cite{CLC_jacs}. The use of QM/MM for this study was mandatory as ion translocation involves proton transfer processes. On the basis of QM/MM MD and well-tempered metadynamics (wtMTD) simulations at the B3LYP and BLYP levels of theory, \citeauthor{CLC_jacs} were able to report proton affinities of the fluoride ion and the gating glutamate residue E148, thus providing valuable insights into transport inhibition. In a second study, the mechanisms of proton transfer and release by the fluoride/proton antiporter CLC\textsuperscript{F}-eca were investigated (Fig. \ref{fig:cases}B)~\cite{CLC-F_jpcl}. Employing the same simulation techniques, it could be shown that a triad is formed between fluoride, glutamate E318, and the gating glutamate E118, eventually releasing protons and fluoride as hydrogen fluoride.

In a recent study, the ligand iperoxo (routinely used in neuroimaging) targeting the human muscarinic acetylcholine receptor 2 was investigated~\cite{Koff_JPCL}. The work focuses on the calculation of the drug unbinding rate constant $k_{\mathrm{off}}$, a very difficult parameter to correctly estimate with force field-based approaches. In fact, while methods based on modern force fields are nowadays able to predict accurate binding free energies and affinities, this is often not the case for rate constants such as $k_{\mathrm{off}}$, which also require a correct estimation of the free energy of transition states. The study shows how sensitive this estimation is to values of the partial charges of the ligand and that while results obtained with a QM/MM MD simulation are in good agreement with experimental findings, standard force field-based procedures lead to qualitatively wrong results, most likely due to the lack of explicit electronic polarization and charge transfer, which are included in QM/MM.

Another important application of QM/MM MD simulations is in scenarios where parts of a system cannot be adequately described using simplified analytical force fields. Notable examples are metal ions, in particular transition metals or divalent alkaline-earth ions. The absence of accurate parameters for the latter hinders a thorough understanding of many biological processes related to, e.g., ion channels, transporters, and pumps. Recently, \citeauthor{Ca_jcim} studied the mechanism of calcium permeation in $\alpha$-amino-3-hydroxy-5-methyl-4-isoxazolepropionic acid receptors (AMPARs), which are key to rapid synaptic transmission in the central nervous system (Fig.~\ref{fig:cases}C)~\cite{Ca_jcim}. In that study, the calcium-binding sites within the channel, initially identified through classical MD, were further confirmed using QM/MM MD. This step was crucial for validating a newly developed classical force field specifically designed for calcium ions. Interestingly, they observed charge transfer between the calcium ion and the water molecules in the first solvation shell, which underlines the importance of a QM/MM description for such systems.

\section{Outlook}

The exascale era is poised to radically change life sciences. With the ability to conduct longer and more accurate simulations on larger and more complex molecular systems than ever before, we will be able to tackle scientific questions that are currently beyond our reach, deepening our understanding of biological processes. This opens up new opportunities for discovery and innovation, potentially revolutionizing fields such as drug design and bioengineering. However, fully realizing the potential of exascale computing is contingent upon overcoming substantial challenges in software design, algorithm optimization, and the efficient utilization of modular and heterogeneous HPC architectures.

The versatility of the MiMiC framework makes it ideal for pushing multiscale biomolecular simulations towards the exascale. It is capable of exploiting the results of the large-scale initiatives by EuroHPC JU and ECP for individual domains such as QM and MM. Indeed, work is already underway to couple a diverse set of QM programs to MiMiC, namely, Quantum ESPRESSO, CP2K, and DFT-FE, all of which are being developed to exploit state-of-the-art HPC technology~\cite{Gavini2023}. Additionally, it is likely that MD simulations will benefit greatly from future ML-based force fields~\cite{Unke2021, Kocer2022}, which MiMiC is fully able to integrate into advanced simulation workflows~\cite{Mouvet2022}. For instance, hybrid ML/MM models facilitate long simulations of biological systems using ML-based force fields, achieving near QM/MM precision at a substantially reduced computational cost~\cite{Fabritiis}, while ML-enhanced free-energy methods significantly accelerate QM/MM calculations of ligand binding affinities for drug discovery~\cite{Rizzi2023}. These and other novel methods are expected to aid in accurately predicting key biophysical properties like drug-protein binding free energies and complete free-energy profiles. We anticipate that the integration of MiMiC-based multiscale simulations with ML and enhanced sampling techniques will further catalyze a \emph{quantum leap} in the fundamental atomistic understanding of biological processes.

\subsection{Acknowledgment}
This preprint was created using the LaPreprint template (\url{https://github.com/roaldarbol/lapreprint}) by Mikkel Roald-Arb\o l \textsuperscript{\orcidlink{0000-0002-9998-0058}}.

\printbibliography

\if@endfloat\clearpage\processdelayedfloats\clearpage\fi






\end{document}